\documentclass[a4paper,12pt]{article}
\usepackage[dvips]{graphicx}
\usepackage{amsmath}
\usepackage[dvips]{color}

\def\10{\bf 10}
\def\5b{\bf {\bar{5}}}

\begin{document}


\vspace{4ex}

\begin{center}
{\large \bf
Spontaneous SUSY breaking with anomalous U(1) symmetry in metastable vacua
and moduli stabilization}

\vspace{6ex}

\renewcommand{\thefootnote}{\alph{footnote}}
S.-G. Kim\footnote{e-mail: sunggi@eken.phys.nagoya-u.ac.jp},
N. Maekawa\footnote{e-mail: maekawa@eken.phys.nagoya-u.ac.jp},
H. Nishino\footnote{e-mail: hiroyuki@eken.phys.nagoya-u.ac.jp}
and 
K. Sakurai\footnote{e-mail: sakurai@eken.phys.nagoya-u.ac.jp}

\vspace{4ex}
{\it Department of Physics, Nagoya University, Nagoya 464-8602, Japan}\\

\end{center}

\renewcommand{\thefootnote}{\arabic{footnote}}
\setcounter{footnote}{0}
\vspace{6ex}


\begin{abstract}

We show that in (anomalous) $U(1)$ gauge theories with the Fayet-Iliopoulos 
(FI) term and with generic interactions there are meta-stable vacua in which
supersymmetry (SUSY) is spontaneously broken even without $U(1)_R$ symmetry
and various hierarchical
structures, for example, Yukawa hierarchy, can be explained by the
smallness of the FI parameter. 
It is shown that adding just one positively charged field to phenomenologically
viable models realizes the spontaneous SUSY breaking.
Moreover, we propose a new scenario for the
stabilization of the moduli in the SUSY breaking models.
It is new feature that the moduli can be stabilized without the superpotential
dependent on the moduli. 
\end{abstract}

\vspace{1cm}


\section{Introduction}
The minimal supersymmetric (SUSY) standard model (MSSM) is one of the
 most promising candidates as the model beyond the standard model 
(SM)\cite{Nilles:1983ge,Haber:1984rc,Martin:1997ns}. It has several attractive features. For example,
 the weak scale can be stabilized
 by the SUSY,
 three gauge couplings meet at a scale which strongly implies the SUSY
 grand unified theory (GUT)\cite{sgut1}-\cite{sgut3}, and the lightest SUSY particle 
 (LSP) can be a dark matter.
 However, there are a lot of unsatisfactory features. One of them is that the
 number of the parameters is 
 more than 100. If we introduce these parameters generically, various
 flavor changing
 neutral current (FCNC) processes 
 and CP violating observables like electric dipole
 moments of electron and neutron become too large  to be consistent
 with the 
  experimental bound 
  \cite{Ellis:1981ts,Barbieri:1981gn,Hagelin:1992tc,Gabbiani:1996hiz}.
   Moreover, it is not known why the supersymmetric Higgs mass
  parameter $\mu$ is of the same order as the SUSY breaking scale
  \cite{Kim:1983dt}.
 Most of these unsatisfied features are strongly related with the SUSY
 breaking. Therefore, it is important to understand the origin of the 
 SUSY breaking in the MSSM in order to solve these problems. Moreover, the large
 hadron collider (LHC) is 
 expected to reveal some features of the SUSY breaking, so it is important to examine various SUSY breaking mechanisms before the LHC gives the results.
 

The (anomalous) $U(1)_A$ gauge theories with the Fayet-Iliopoulos(FI) 
term \cite{xi} are
often used
to explain the hierarchical structures of Yukawa
couplings \cite{yukawa1}-\cite{yukawa6}. It is quite reasonable because the hierarchical structures 
can be explained under the assumption that all the interactions which  
 are allowed by the symmetry are introduced with O(1) coefficients.
 Moreover, it has been pointed out that the $U(1)_A$ symmetry can play an 
 important role even in breaking grand unified group
 \cite{yukawa6}-\cite{Chkareuli:2000bm}.
 This is also 
 natural because the serious fine-tuning problem called the doublet-triplet splitting
 problem can be solved under the same assumption in which the  generic interactions
 are introduced \cite{yukawa6, M1}.
 
 In the literature, it has been argued that even SUSY can be spontaneously 
 broken with the (anomalous) $U(1)_A$ symmetry with the FI-term \cite{po}. In order to
 break SUSY with generic interactions, the $U(1)_R$ symmetry must be
 imposed
 \cite{Nelson:1993nf}.
 In other words, without $U(1)_R$ symmetry, 
 SUSY vacua appear in general.   
However, the above phenomenological models have often no
 $U(1)_R$ symmetry. Therefore, it is important to examine the spontaneous 
 SUSY breaking without $U(1)_R$ symmetry. 
This may be possible, if we consider the meta-stable vacua
 \cite{meta1, meta2, Amariti:2006vk, Dienes:2008gj}
\footnote{In Ref.\cite{Dienes:2008gj}, the meta-stable SUSY breaking with FI-term is discussed, 
though in their model, $U(1)_R$ symmetry is imposed.}
.
In this paper, we point out that even if generic interactions are
 introduced in (anomalous) $U(1)_A$ gauge theory with the FI-term and without 
 $U(1)_R$ symmetry, SUSY can be spontaneously
   broken in meta-stable vacua,
  in which various hierarchical parameters are determined by the 
  smallness of the FI parameter. If generic interactions are introduced
  with O(1) coefficients, almost all the scales can be determined by the 
  symmetry of the theory (i.e., the $U(1)_A$ charges). We calculate the
  various scales including several SUSY breaking scales in some examples. 
One of the most interesting features in the meta-stable spontaneous SUSY breaking 
proposed in this paper is that adding just
one positively charged field to phenomenologically viable models mensioned in the
previous paragraph
 realizes the 
spontaneous SUSY breaking.
This makes us expect more complete models in which in addition to the previous
advantages of the models with anomalous $U(1)_A$ symmetry, SUSY breaking is also controlled by the anomalous $U(1)_A$ gauge symmetry.

One of the most important problem in the phenomenology of the superstring theory 
is the moduli stabilization problem\cite{Dine:1985he}, though there are several
attempts to solve this problem in various scenarios in which SUSY is dynamically broken
by the strong dynamics of supersymmetric QCD (SQCD)
\cite{Krasnikov:1987jj}-\cite{ArkaniHamed:1998nu}.
 Especially in the models
with the anomalous $U(1)_A$ symmetry, the gauge anomaly is cancelled by the
shift of the moduli, and in general, the FI parameter is determined by the 
VEV of the moduli
\cite{Witten:1981kv,Dine:1987xk,Atick:1987gy,Dine:1987gj}.
 Therefore, in the context of the SUSY breaking models with
the anomalous $U(1)_A$ symmetry, it is an interesting and challenging subject 
to consider the moduli stabilization simultaneously. We propose a possibility
to stabilize the moduli in the SUSY breaking scenario. As the result, we can
obtain a SUSY breaking scenario in which SUSY is spontaneously broken and the
moduli can be stabilized without the superpotential dependent on the moduli. 
In the literature, 
the superpotential dependent on the moduli,
which is induced
by non-perturbative effect or by the flux compactification \cite{Giddings:2001yu},
 plays
an essential role in stablizing the moduli. However, in the moduli stabilization
mechanism proposed in this paper, the superpotential does not include the 
moduli\footnote{
Generically, the symmetry allows the exponential type interactions of the moduli
and the inverse power type interactions of the introduced fields, which may be
induced by the non-perturbative effects of the strong dynamics or of the string.
In this paper, we consider the case in which such interactions are not induced or
are sufficiently small if any, because these interactions spoil the SUSY zero 
mechanism which plays an important role in solving various phenomenological
problems. Such an assumption may be reasonable in our setup 
because in order to break SUSY we do not require any strong coupling
gauge theory which has the dynamical scale larger than the weak scale.
}.
This feature is quite important because the superpotential dependent on the moduli
generically spoils the SUSY zero mechanism \cite{yukawa6}\cite{M1}\cite{Nir:1993mx}
which plays an important role in building realistic
models.

Organization of this paper is as follows.
In the second section, we will compare two simple 
spontaneous SUSY breaking models with the anomalous $U(1)_A$ symmetry.
The $U(1)_R$ symmetry is imposed in one model and not in the other.
The latter model has meta-stable SUSY breaking vacua in which the 
hierarchical couplings as Yukawa couplings can be realized.
In the third section, we extend the meta-stable SUSY breaking model
without $U(1)_R$ symmetry to the more general models.
Applying this general results to phenomenologically viable models, 
it is easily understood that adding one positively charged field to
the models realizes
spontaneous SUSY breaking.
And in the fourth section, we will consider the moduli stabilization.
Finally, we will give a summary and discussions.

\section{Spontaneous SUSY breaking with anomalous $U(1)_A$ symmetry}
In this section, we consider SUSY breaking models with anomalous $U(1)_A$ gauge symmetry. And we show that there are meta-stable SUSY breaking vacua 
in a simple model with generic interactions without $U(1)_R$ symmetry.
In the vacua, hierarchical couplings can be obtained.

Before we consider the  SUSY breaking model without $U(1)_R$ symmetry, let us 
recall what happens with $U(1)_R$ symmetry \cite{xi}.
For simplicity, we consider a model which contains two fields $S$ and $\Theta$, where $S$ has positive integer $U(1)_A$ charge $s$ and $\Theta$ has 
negative $U(1)_A$ charge $\theta=-1$. (In this paper, we use the lowercase letter
as the charge for the field denoted by the uppercase letter.)
We assign R charges for $S$ and $\Theta$ 
 as in table 1.
\begin{table}[h]
\begin{center}
\begin{tabular}{c|ccc}
~& $ W $& $S$ & $\Theta$ \\ \hline
$ U(1)_A $ & $0$& $s>0$ & $-1$ \\ 
$ U(1)_R $& $2$ & $ 2 $& $0$  
\end{tabular}
\end{center}
\caption{The quantum numbers of the superpotential $W$ and the fields, $S$ and $\Theta$.}
\end{table}
In this model, the generic superpotential becomes
\begin{equation}
W = S \Theta^s 
\label{eqn:51}
\end{equation}
where the coefficients are neglected. (In this paper, we usually neglect the 
coefficients in the interactions and take 
the cutoff $\Lambda=1$.)
The $F$-terms and the $D$-term in this model,
\begin{equation}
\begin{aligned}
F^*_S &=-\frac{\partial W}{\partial S} = -\Theta^s, \\
F^*_{\Theta} &=-\frac{\partial W}{\partial \Theta} = -s S \Theta^{s-1}, \\
D_A &= -g \Big(\xi^2 -|\Theta|^2 + s|S|^2  \Big),
\label{eqn:52}
\end{aligned}
\end{equation}
where $\xi$ is a FI parameter,
cannot be vanishing simultaneously because the F-flatness conditions
results in the vanishing VEV of $\Theta$ under which the D-flatness
condition cannot be satisfied.
Therefore SUSY is spontaneously broken in this model.
The VEVs of these fields, $F$ and $D$ terms are determined by the minimization
of the potential
\begin{equation}
V=|F_S|^2+|F_\Theta|^2+\frac{1}{2}D_A^2
\end{equation}
as
\begin{eqnarray}
\langle S\rangle&=&0,\quad
\langle \Theta\rangle=\lambda\\
\langle F_S\rangle&\sim&\lambda^s,\quad
\langle F_\Theta\rangle=0,\quad
\langle D_A\rangle\sim\frac{s}{g}\lambda^{2s-2},
\end{eqnarray}
when $\xi\ll 1$. Here, $\lambda\equiv\langle\Theta\rangle/\Lambda\sim\xi/\Lambda$,
and without loss of generality, we can take 
 the VEV of $\Theta$ real because of the $U(1)_A$ symmetry.
The typical SUSY breaking scale, $\lambda^s\Lambda$, must be 
around the weak scale, which is obtained, for example, when 
$s\sim 24$ for $\lambda\sim 0.22$ and 
$\Lambda=2\times 10^{18}$ GeV.


What happens if we do not impose $U(1)_R$ symmetry? 
The quantum numbers are given as in table \ref{table2}.
\begin{table}[h]
\begin{center}
\begin{tabular}{c|ccc}
~& $S$ & $\Theta$ \\ \hline
$ U(1)_A $ & $s>0$ & $-1$  
\end{tabular}
\end{center}
\label{table2}
\caption{The $U(1)_A$ charges of the fields, $S$ and $\Theta$}
\end{table}
Then, the generic superpotential becomes
\begin{equation}
W(S\Theta^s) = \sum_{n=1} a_n\left( S \Theta^s \right)^n. 
\label{eqn:512}
\end{equation}
Namely, any polynomial of $x\equiv S\Theta^s$ is allowed for the superpotential
$W(x)$.
It is known that such a model has SUSY vacua. 
Actually, among the $F$-term and the $D$-term 
\begin{equation}
\begin{aligned}
F^*_S &=-\frac{\partial W}{\partial S} = -\frac{\partial W}{\partial x}\Theta^s \\
F^*_{\Theta} &=-\frac{\partial W}{\partial \Theta} =-\frac{\partial W}{\partial x}
 s\Theta^{s-1} S \\
D_A &= -g \Big(\xi^2 -|\Theta|^2 + s|S|^2  \Big),
\label{eqn:513}
\end{aligned}
\end{equation}
the $F$-flatness conditions can be satisfied by taking the VEV of $S\Theta^s$ 
to be $\partial W/\partial x=0$ and the $D$-flatness can be satisfied by choosing the VEV of 
$\Theta$. Generically the VEVs of $S$ and $\Theta$ become $O(1)$.
However, it has not been emphasized that this model has meta-stable SUSY
 breaking vacuum at $\langle{\Theta}\rangle\sim\xi$ and 
 $\langle{S}\rangle\sim 0$ if $\xi\ll 1$. Note that such VEVs play an
 important role in solving various phenomenological 
 problems\cite{yukawa1}-\cite{M1}.
The reason for the meta-stability is, roughly speaking, that
in the region $\langle{S}\rangle, \langle{\Theta}\rangle\ll 1$,
the superpotential becomes $W=S\Theta^s$ approximately which is nothing
but the superpotential in the spontaneous SUSY breaking model with $U(1)_R$ symmetry.

In order to estimate the VEVs $\langle S\rangle=S_re^{i\phi_s}$, 
$\langle\Theta\rangle=\Theta$, $\langle F_S\rangle$, $\langle F_\Theta\rangle$, 
and $\langle D_A\rangle$ and see the meta-stability of this vacuum,
we must examine the potential
\begin{equation}
V=\left|\frac{\partial W}{\partial x}\right|^2
(\Theta^{2s}+s^2S_r^2\Theta^{2(s-1)})+\frac{g^2}{2}(\xi^2-\Theta^2+sS_r^2)^2,
\label{eqn:01}
\end{equation}
where $\frac{\partial W}{\partial x}=\sum_{n=1} a_nn(S_r e^{i\phi_s}\Theta^s)^{n-1}$.
Suppose the small deviation from the VEVs,
%
%
$\left<S \right>=0$ and 
$ \left<\Theta \right>=\xi \ll 1$.
Then, it is sufficient to examine the superpotential up to the second order as
$W=a_1(S\Theta^s)+a_2(S\Theta^s)^2$ because of the smallness of the VEVs of $S$ and
$\Theta^s$. The stationary conditions
\begin{equation}
\frac{\partial V}{\partial \Theta  }=0,\quad
\frac{\partial V}{\partial S_r}=0,\quad
\frac{\partial V}{\partial \phi_s}=0
\end{equation}
lead to 
\begin{eqnarray}
&&\langle D_A\rangle\sim \frac{s|a_1|^2}{g}\Theta^{2s-2}\sim 
\frac{s}{g}\lambda^{2s-2}, \label{dflat} \\
&&S_r\sim
 -\frac{\Theta^{s+2}}{2s^2|a_1|^2}(a_2a_1^*e^{i\phi_s}+h.c.)\sim\frac{1}{s^2}
 \lambda^{s+2},\\
&&a_1^*a_2e^{i\phi_s}-a_1a_2^*e^{-i\phi_s}=0. 
\end{eqnarray}
If we define $\delta \Theta = \Theta - \xi$,
eq.\,(\ref{dflat}) implies $\delta \Theta \sim -\frac{s|a_1|^2}{2g} \xi^{2s-3}$.
This is consistent with our assumption $\delta \Theta / 
\xi \ll 1$.
Then, the VEV of auxiliary fields are determined as
$-F_S^* \sim a_1 \lambda^s$ and $-F_\Theta^*\sim \frac{\lambda^{2s+1}A}{sa_1^*}$, 
where $A\equiv -a_1a_2^*e^{-i\phi_s}=-a_1^*a_2e^{i\phi_s}$. 
Moreover, the stability condition requires $a_1^*a_2e^{i\phi_s}=-|a_1a_2|$.
One of the biggest differences between the above two SUSY breaking models is 
the value of the VEV of the positively charged field $S$. (It is obvious that the 
difference of the VEV of $F_\Theta$ is caused by the difference of the VEV of $S$.)
Therefore, we will briefly examine the reason for the difference below.
In the model with $U(1)_R$ symmetry
the VEV of $S$ is vanishing, while in the model without $U(1)_R$ symmetry, $S$ has
non-vanishing VEV, though the VEV is smaller
than the typical SUSY breaking scale $F_S/\Lambda\sim\lambda^s$.
Without $U(1)_R$ symmetry, the superpotential includes higher dimensional operators
like $S^2\Theta^{2s}$. The term leads to a tadpole term 
$\langle F_S\rangle \lambda^{2s}S$ after obtaining
non-vanishing VEV of $F_S$, which results in the non-vanishing VEV of $S$.
Namely, the VEV $ \left< S \right>$ is decided by the tadpole and the mass term 
$\lambda^{s-1}S\Theta$ as
\begin{equation}
\begin{aligned}
\left< S \right>\sim\frac{\text{coefficient of tadpole}}{\text{mass}^2}
\label{eqn:519}
\end{aligned}
\end{equation}\\
as seen in Figure 1.
\begin{figure}[h]
\begin{center}
\includegraphics[width=5cm,clip]{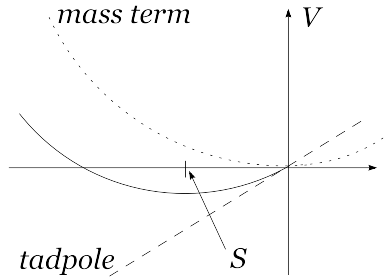}
\end{center}
\caption{This figure shows the potential of $S$. 
The VEV of $S$ is decided by the tadpole and the mass term of $S$.
The solid line shows sum of the contributions from the
tadpole and mass terms.}  
\end{figure}
It is obvious that the larger tadpole term leads to lower potential energy.
Therefore, the phase of the VEV of $S$ is determined so that the tadpole term
$W|_{\theta^2}+h.c. \ni SS\Theta^{2s} |_{\theta^2}+h.c.\sim S F_S \lambda^{2s}+h.c.=2\lambda^{3s}S\cos\phi_s $ becomes maximal, i.e., $\cos\phi_s=\pm 1$. 
(Here we take the coefficients in the superpotential real for the simplicity.)
The signature of $\cos\phi_s$ is determined so that the absolute value of 
$\partial W/\partial x=a_1+a_2S_re^{i\phi_s}\Theta^{s}$  is minimal, i.e., the signature becomes
minus if $a_1$ has the same signature as $a_2$. (Here we use the notation
$S$ and $\Theta$ are positive.)
Since the mass term is given by $ 
\left| \frac{ \partial W }{ \partial \Theta}
 \right|^2 \ni \lambda^{2s-2}|S|^2  $, 
the VEV $\left< S \right>$ become $ \left<S \right>\sim \lambda^{s+2}$ from eq. $(\ref{eqn:519})$. 
If $\lambda^s \Lambda $ is the weak scale and the cut off $\Lambda$ is much larger than the weak scale, 
the VEV $ \left<S \right>$ would be much smaller than the cut off $\Lambda$ and the 
VEV $\left< \Theta  \right>$. 
As the results, these values of VEVs are approximately satisfied with the expected
VEV relations which are important in solving 
phenomenological problems.

We show the schematic form of the potential in this model in the Figures 1 and 2. 
The potential of $\Theta$ rapidly increase above $\Theta=1$, 
because of $|F_S|^2$. The Figure 3 shows the magnification of the potential around the origin.

In the last part of this section, we estimate the lifetime of the meta-stable vacuum
by following the arguments in the references 
\cite{lifemeta} with the values $s=24$ for $\lambda=0.22$. 
Lifetime of the meta-stable vacuum is approximately given by
\begin{equation}
\begin{aligned}
\tau \propto e^P,
\label{eqn:521}
\end{aligned}
\end{equation}
where $P$ is dimensionless and can be given by
\begin{equation}
\begin{aligned}
P=\frac{( \sqrt{V_h}  \Delta )^4}{\epsilon '^3}.
\label{eqn:522}
\end{aligned}
\end{equation}
$\Delta$ shows the distance from supersymmetric vacuum to meta-stable vacuum. 
$V_h$ shows the height of the barrier wall between the meta-stable vacuum and 
SUSY vacua
and $\epsilon '$ shows the potential height of the meta-stable vacuum.
This model become $\Delta \sim O(\Lambda),V_h \sim O(\Lambda^4),
\epsilon ' \sim O(\lambda^{2s}\Lambda^4)\sim O(M_{SUSY}^2\Lambda^2)$ ,
where $M_{SUSY}$ is the SUSY breaking scale.
Therefore,  we can estimate $P$ as
\begin{equation}
P \sim \frac{\Lambda^{12}}{M_{SUSY}^6\Lambda^6} \sim \frac{\Lambda^6}{M_{SUSY}^6}.
\label{eqn:523}
\end{equation}
Then we obtain 
$P \sim 10^{96}$. Therefore, the lifetime of the meta-stable vacuum is
much larger than the age of our
universe.
\begin{figure}[h!]
\hspace{-4cm}
\begin{center}
\includegraphics[width=8cm,clip]{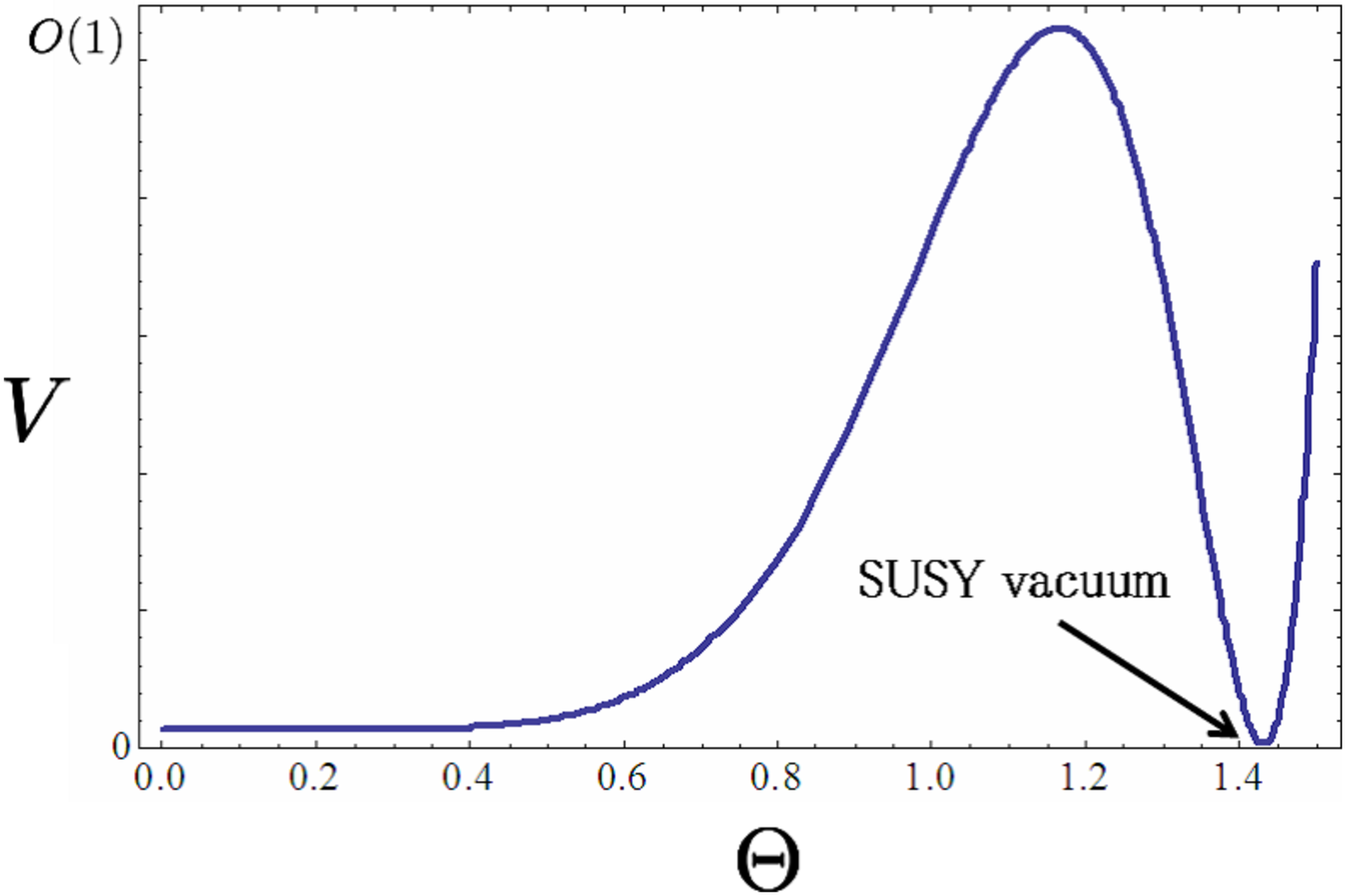}
\end{center}
\caption{
This figure shows the potential of $\Theta$. 
Here, we take $s=4$ and $\xi=0.2$.
}  
\label{non-RRTT3}
\begin{center}
\includegraphics[width=8cm,clip]{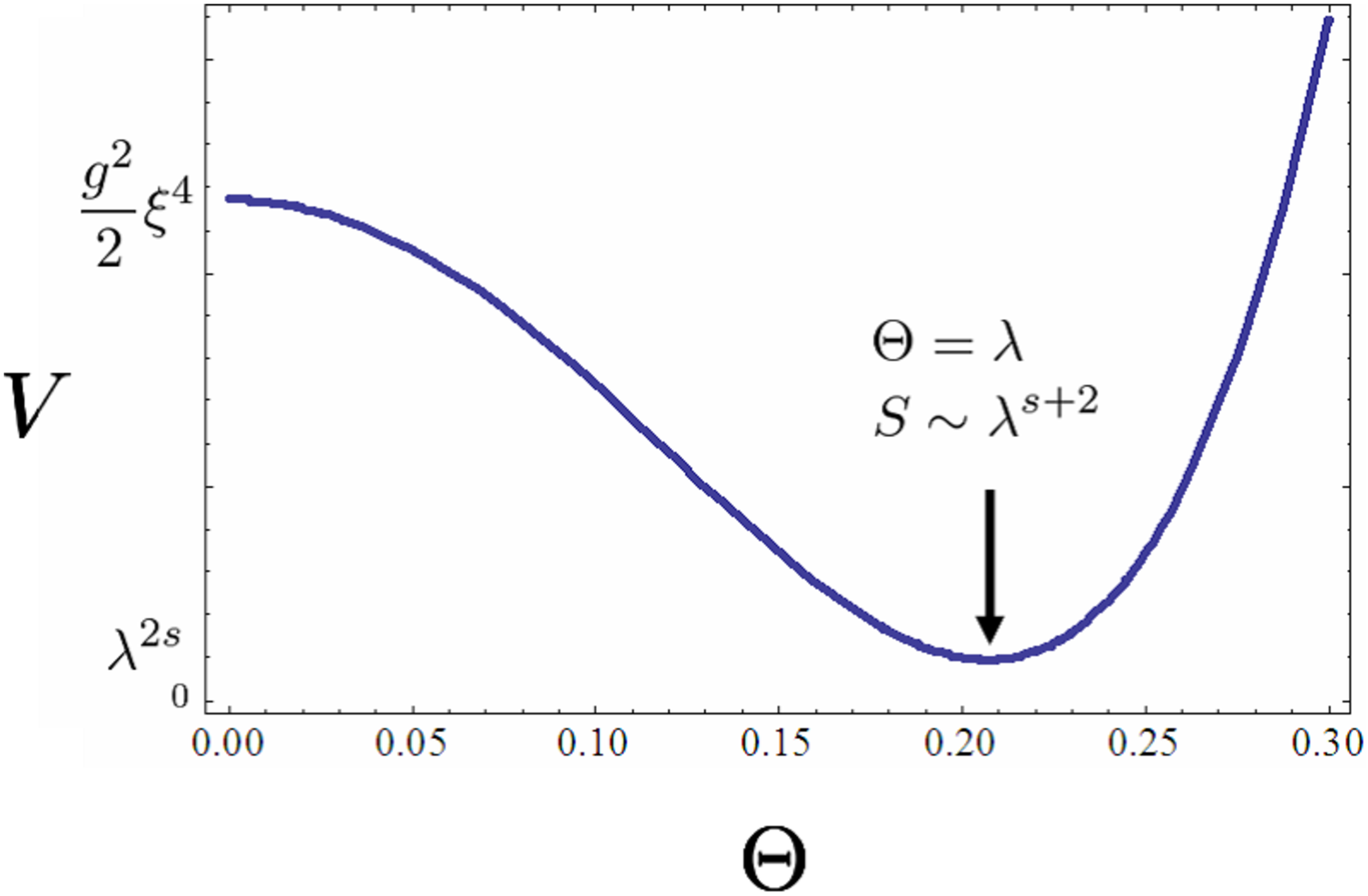}
\end{center}
\caption{This figure shows the magnification of the potential 
in Fig.~(\ref{non-RRTT3}) around the origin.}
\label{non-RRTT2}
\end{figure}

\section{General cases}
In this section, we will extend the SUSY breaking model discussed in the previous 
section to more general ones.
We introduce $n_+$ positively charged fields $S_i (i=1,\cdots, n_+)$ and
$n_-$ negatively charged fields $Z_j (j=1,\cdots, n_--1)$ and $\Theta$
as in table 3.
\begin{table}[h]
\begin{center}
\begin{tabular}{c|ccc}
~& $ S_i $ & $Z_j$ & $\Theta$ \\ \hline 
$ U(1)_A $& $s_i>0$ & $z_j<-1$ & $-1$\\ 
\end{tabular}~~
$\big( i=1 \sim n_+~,~j=1 \sim (n_- -1) \big)$
\end{center}
\caption{The $U(1)_A$ charges for the fields $S_i$, $Z_j$, and $\Theta$.}
\end{table}\\
One of the $n_++n_-$ complex F-flatness conditions becomes trivial because of the 
$U(1)_A$ gauge symmetry, but we have the real D-flatness condition for the $U(1)_A$.
Then the VEVs of $n_++n_-$ complex fields are generically fixed by these conditions
except one real field which corresponds to the Nambu-Goldstone mode of the $U(1)_A$
symmetry if all the conditions are independent.
Therefore, if we introduce the generic superpotential, $W(S,Z,\Theta)$, namely, 
the above 
conditions become independent, then, in general, there are
SUSY vacua at which all the VEVs of the fields are of order one if all the 
coefficients are of order one.
However, as discussed in Refs.\cite{yukawa6,M1}, when $n_+\leq n_--1$, other SUSY vacua appear at
which all positively charged fields, $S_i$, have vanishing VEVs and the negatively
charged fields, $Z_j$ and $\Theta$, 
have non-vanishing VEVs which are not larger than O($\xi$)$\ll 1$.
When all positively charged fields have vanishing VEVs, the F-flatness conditions
of negatively charged fields are trivially satisfied because 
$\partial W/\partial Z_j$ have positive charges. Therefore, the $n_+$ 
F-flatness conditions and the D-flatness condition of $U(1)_A$,
\begin{equation}
\frac{\partial W}{\partial S_i}=0,\quad D_A=g(\xi^2-|\Theta|^2+\sum_j z_j|Z_j|^2)=0,
\end{equation}
constrain the
$n_-$ VEVs of negatively charged fields, $Z_j$ and $\Theta$. If $n_+\leq n_--1$,
these conditions can be satisfied in general, and therefore, there are SUSY vacua.
Because of the D-flatness conditions, the non-vanishing VEVs cannot be larger than
$\xi\ll 1$. (In this paper, we call such vacua small vacua.)
Especially, when $n_+=n_--1$, then all the VEVs are determined by their charges as
\begin{equation}
\langle S_i\rangle=0,\quad \langle Z_j \rangle\sim\lambda^{-z_j}.
\end{equation}
Since the generic superpotential can be rewritten as
$W(\tilde S_i, \tilde Z_j)$, where $\tilde S_i=S_i\Theta^{s_i}$ and 
$\tilde Z_j=Z_j\Theta^{z_j}$, the F-flatness conditions of $S_i$
\begin{equation}
\frac{\partial W}{\partial S_i}=\Theta^{s_i}\frac{\partial W}{\partial \tilde S_i}
\label{Fs}
\end{equation}
give solutions as $\langle \tilde Z_j \rangle=O(1)$ because 
$W(\tilde S_i, \tilde Z_j)$ have $O(1)$ coefficients. 
The equations $\langle \tilde Z_j\rangle=O(1)$ mean 
$\langle Z_j \rangle\sim \lambda^{-z_j}$
\footnote{Of course, even if $n_+\leq n_--1$, it can happen that there
is no such vacua, i.e., under the assumption that all the positively charged
fields have vanishing VEVs, all the F and D flatness conditions cannot be satisfied.
For example, if one positive charge $s_1$ is smaller than all the magnitudes of
the negative charges $z_j$, then the F-flatness condition of $S_1$ and the 
D-flatness condition cannot be satisfied simultaneously. Here, we do not consider
such extreme cases.}. Usually, this is the case in the most of 
phenomenologically viable models.

What happens if $n_+>n_--1$? Since the number of the constraints is larger than the
 number of the variables, there is no solution, and therefore, small vacua cannot
 be in the supersymmetric vacua
as discussed in Refs. \cite{yukawa6,M1}.
However, as discussed in the previous section, the small vacua can be meta-stable.
Let us figure out what happens if $n_+=n_-$. One of the F- and D-flatness
conditions cannot be satisfied. If the F-flatness condition of the largest charged
field $S_{n_+}$ is not satisfied and the other F and D flatness conditions are
almost satisfied, the vacuum energy becomes the lowest in the small vacua
because the eq. (\ref{Fs}) gives $|F_{S_i}|\sim \lambda^{s_i}$.
Therefore, the vacuum energy becomes $V\sim |F_{S_{n+}}|^2\sim\lambda^{2s_{n_+}}$,
which can be very small if the maximal charge $s_{n_+}\gg 1$. This feature
may give an explanation of the large hierarchy between the SUSY breaking scale
and the Planck scale.
It is reasonable to expect that the potential energy becomes larger than 
$\lambda^{2s_{n_+}}$ between the small vacua and the SUSY vacua
at which all the VEVs are of order one. When the VEVs become larger
than $\xi$, the D-flatness condition requires the non-vanishing VEVs
for positively charged fields. Then all the $F$ terms including those of negatively
charged fields can contribute to the vacuum energy
 which are generically become larger than $\lambda^{2s_{n_+}}$.
Note that if we add one positively charged field to the phenomenologically viable
model in which $n_+=n_--1$, we can obtain the model in which SUSY is spontaneously
broken by the meta-stable vacua.

\section{Moduli stabilization in a model with anomalous $U(1)_A$ symmetry }
In the previous sections, we have assumed that the FI parameter $\xi$ is a constant.
However, in the context of the supergravity or the superstring, the FI parameter
is dynamically determined, i.e., $\xi$ depends on the VEV of the moduli (or dilaton) 
$D$ \cite{Witten:1981kv,Dine:1987xk,Atick:1987gy,Dine:1987gj}. Actually, since the $U(1)_A$ gauge symmetry is given by
\begin{eqnarray}
V_A&\rightarrow& V_A+\frac{i}{2}(\Lambda-\Lambda^\dagger)\\
D&\rightarrow&D+\frac{i}{2}\delta_{GS}\Lambda,
\end{eqnarray}
where $\Lambda$ is a parameter chiral superfield. 
A dimensionless parameter $\delta_{GS}$, which is proportional to ${\rm tr}Q_A$,  
is positive when
${\rm tr} Q_A>0$.
The K\"ahler potential $K_D(D+D^\dagger-\delta_{GS}V_A)$ is invariant under
$U(1)_A$ gauge symmetry and the FI term can be given as
\begin{equation}
\int d^4\theta K_D(D+D^\dagger-\delta_{GS}V_A) =
-\left(\frac{\delta_{GS}K_D'}{2}\right)D_A+\cdots\equiv\xi^2D_A+\cdots,
\label{xi}
\end{equation}
where we take the sign of the trace of anomalous $U(1)_A$ charge tr$Q_A$ so that
$\xi^2>0$. (Since tr $Q_A>0$ in the most of phenomenologically viable models, 
the positivity of FI parameter requires $K'_D<0$, which is consistent with the 
stringy tree level K\"ahler potential of the moduli, 
$K_D=-\ln (D+D^\dagger-\delta_{GS}V_A)$.)

The stabilization of the moduli is one of the important issues in the 
string theory and/or in the models with the anomalous $U(1)_A$ gauge symmetry.
In this section, we will examine a new possibility for the moduli stabilization
by using the potential dependent on the moduli through the FI parameter $\xi$,
which is obtained in the previous sections as
\begin{equation}
\begin{aligned}
V \sim |F_S|^2 \sim \xi^{2s}.
\label{eqn:dilaton2}
\end{aligned}
\end{equation}
First, we examine the stabilization by the deformation of the K\"ahler potential
of the moduli $K_D$ from 
$K_D=-\ln(D+D^\dagger-\delta_{GS}V_A)$ which can be obtained by stringy calculation
at tree level. However, unfortunately we found it impossible. 
The point is simple.
It is shown that $ \xi^{2s} (D) $ is monotonically decreasing function for $D$.
Actually, 
\begin{equation}
\begin{aligned}
\frac{\partial\xi^{2s}}{\partial D}=\big( \xi^{2s} \big)'=sK''_D(K'_D)^{s-1}(-\frac{\delta_{GS}}{2})^s<0, 
\label{eqn:630}
\end{aligned}
\end{equation}
where $K''_D$ is positive because it becomes
 the coefficient of the moduli kinetic term. 
%
%
This result shows it difficult to stabilize the moduli by the deformation of 
$K_D$. 
%

%
Next, we will consider the deformation of the K\"ahler potential of $S$ from the
canonical form.
%
Since the scalar potential of the moduli can be obtained as
\begin{equation}
\begin{aligned}
V & \sim \left ( \frac{\partial ^2 K_S}{\partial S \partial S^{\dag}} \right )^{-1}
 \left | \frac{\partial W}{\partial S} \right |^2 
& \sim \left ( \frac{\partial ^2 K_S}{\partial S \partial S^{\dag}} \right )^{-1} \xi^{2s} (D),
\label{eqn:D12}
\end{aligned}
\end{equation}
the moduli can be stabilized as in Fig. 4 
if $ \frac{ \partial ^2 K_S }{ \partial S \partial S^{\dag} } $ becomes much smaller than one at $\langle D\rangle=D_0$.

\begin{figure}[h]
\begin{center}
\includegraphics[width=8.5cm,clip]{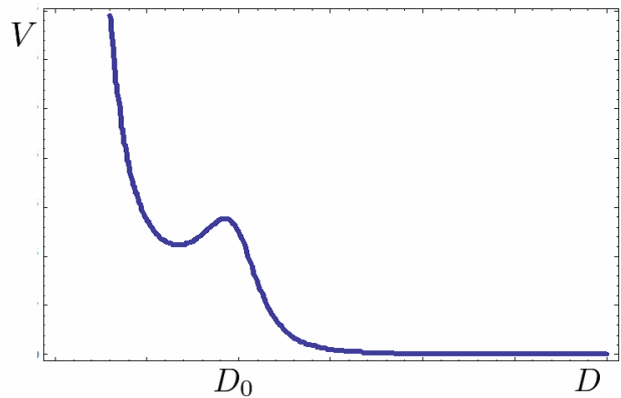}
\end{center}
\caption{The potential of the moduli.}  
\end{figure}
In order to realize such a situation, let us take  
more generic K\"ahler potential of the $S$ field
as
\begin{equation}
\begin{aligned}
K_S = S^{\dag} Sf(D+D^{\dag} -\delta_{GS} V_A),
\label{eqn:D11}
\end{aligned}
\end{equation}
where $f(x)$ is a function of $x$. If the function $f(x)$ is given by
\begin{equation}
\begin{aligned}
f(x)=c(x-x_0)^2 + \epsilon,
\label{eqn:D13}
\end{aligned}
\end{equation}
where 
$c\sim O(1),~0<\epsilon \ll 1 $, 
then the function
$ \frac{\partial ^2 K_S}{\partial S \partial S^{\dag}}  $ becomes much smaller than one at $x=x_0$. 
The moduli potential (\ref{eqn:D12}) can be rewritten by using the K\"ahler 
potential $(\ref{eqn:D11})$ as 
\begin{equation}
V \sim \left ( \frac{\partial ^2 K_S}{\partial S \partial S^{\dag}} \right )^{-1}
 \left | \frac{\partial W}{\partial S} \right |^2 
\sim \frac{\xi^{2s}(x)}{\left(c(x-x_0)^2 + \epsilon \right)}
\sim \frac{1}{\left(c(x-x_0)^2 + \epsilon \right)}\left(\frac{\delta_{GS}}
{2x}\right)^s,
\label{eqn:D14}
\end{equation}
where the last equality is obtained from the eq. (\ref{xi}) and 
$K_D(x)\sim -\ln x$.
It is easily shown that if the condition 
\begin{equation}
\epsilon < \frac{c}{s(s+2)} x^2_0
\label{eqn:D19}
\end{equation}
is satisfied, the moduli potential has the local minimum at $x=x_-$ and
the local maximum at $x=x_+$ as in Fig. 4, where 
\begin{equation}
x_\pm\equiv \Big(\frac{s+1}{s+2}\Big)x_0\left \{1\pm\sqrt{1 - \gamma } \right \},
\quad 
\gamma \equiv \frac{s(s+2)}{(s+1)^2} \Big( 1 +\frac{ \epsilon }{c x^2_0} \Big).
\end{equation}
%

Let us estimate the scales of 
$ F_S \sim - \frac{ \frac{\partial W}{{\partial S}}}{\frac{\partial ^2 K_S}{\partial S \partial S^{\dag}}}$
and $D_A$. 
At the meta-stable 
vacuum $x=x_-$, $\frac{\partial ^2 K_S(x)}{\partial S \partial S^{\dag}}$ can be estimated as
\begin{eqnarray}
\left. \frac{\partial ^2 K_S(x)}{\partial S \partial S^{\dag}} \right |_{(x=x_-)}
&=&\frac{cx_0^2}{(s+2)^2}\left(2-\frac{s(s+2)\epsilon}{cx_0^2}
+2\sqrt{1-\frac{s(s+2)\epsilon}{cx_0^2}}\right) \\
&\sim&\left\{\begin{array}{ll}
                                    \frac{4}{(s+2)^2}cx_0^2,\quad 
                                    &(\epsilon\ll\frac{cx_0^2}{s(s+2)})\\
                                    \frac{1}{(s+2)^2}cx_0^2, \quad 
                                    &(\epsilon\sim\frac{cx_0^2}{s(s+2)}).
                                  \end{array}\right.
\end{eqnarray}
Namely,  $ \frac{\partial ^2 K_S}{\partial S \partial S^{\dag}} \sim\frac{x_0^2}{s^2}\ll 1$ for $s\gg 1$ and $c\sim 1$.
Therefore, $F_S\sim s^2\lambda^s/x_0^2$.
>From the scalar potential
\begin{equation}
V\sim \frac{\partial ^2 K_S}{\partial S \partial S^{\dag}} |F_S|^2+\frac{1}{2}D_A^2,
\end{equation}
we obtain
\begin{equation}
D_A\sim \frac{s^3}{x_0^2}\lambda^{2s-2}.
\end{equation}
In the previous sections, we obtained
$D_A\sim s\lambda^{2s-2}$ which is much larger than $|F_S|^2\sim\lambda^{2s}$.
In the scenario in which the FI term is dynamically determined, the ratio
$D_A/|F_S|^2$ becomes smaller.

Let us examine the concrete values of the parameters.
To obtain $F_S\sim O(100{\rm GeV})$, $s\sim 28$ is required for $\lambda=0.2$,
$\Lambda=10^{18}$ GeV, $c=1$, and $x_0=1$.
Then, to satisfy the condition 
eq. (\ref{eqn:D19}), the parameter $\epsilon$ must be smaller than $10^{-3}$. 
The ratio $D_A/|F_S|^2$ becomes of order one.
This may be important in applying this mechanism of SUSY breaking
to the realistic SUSY breaking models.

The lifetime of this meta-stable vacuum can easily be longer than the age of the 
universe. 
Let us estimate the lifetime of the meta-stable vacuum
by using $(\ref{eqn:521}),(\ref{eqn:522})$
with $V_h=V(x_+)-V(x_-) $, $\epsilon '=V(x_-)$, and 
$\Delta=x_+-x_-$ which is taken for conservative estimation.
If we take the parameters as $s=28~,c=1~,\epsilon =10^{-3}~$, and $x_0=1~$,
which satisfy the condition $(\ref{eqn:D19})$, 
then $P$ can be estimated as
\begin{equation}
P >\frac{ \big( \sqrt{V_h} \Delta \big)^4 }{(\epsilon ')^3} 
 \sim 10^{31}. 
\label{eqn:D21}
\end{equation}
Therefore, the lifetime of the meta-stable vacuum becomes much longer than  
the age of the universe.

Note that we do not use extra SQCD dynamics to 
break SUSY and/or stabilize the moduli. 
In this scenario, the SUSY breaking scale which is much smaller than the Planck
scale is obtained by the smallness of the FI parameter and the large anomalous 
$U(1)_A$ charge of the $S$ field.
Therefore, this new scenario for the spontaneous SUSY breaking is economical.
This is one of the most crucial difference between this scenario and the previously
proposed scenarios\cite{Krasnikov:1987jj}-\cite{ArkaniHamed:1998nu}.

\section{Discussion and conclusion}
We proposed a new SUSY breaking scenario which can be applied to the most of the
phenomenologically viable models with anomalous $U(1)_A$ gauge symmetry.
Even without the $U(1)_R$ symmetry, which usually plays an essential role in breaking
SUSY spontaneously, SUSY can be broken spontaneously, because the SUSY breaking
vacua are meta-stable. Moreover, we examined the moduli stabilization in the 
scenario. And we found that the stabilization is possible by the deformation
of the K\"ahler potential though some tuning of parameters is required.
It is important that the stabilization of the moduli can be realized
without the superpotential dependent on the moduli, because such superpotential
generically spoils the SUSY zero mechanism which plays an critical role in obtaining
phenomenologically viable models.



One of the easiest application of this SUSY breaking scenario is that
SUSY is spontaneously broken in the hidden sector by this scenario instead
of the dynamical SUSY breaking scenario. 
Another interesting and important subject is to examine the possibility
that this SUSY breaking mechanism is applied in the visible sector and 
the realistic mass spectrum of superpartners of the standard model particles
is obtained at the same time, i.e., the hidden sector is not needed.
Unfortunately, we have several obstacles for this subject. 
One of the most serious issue is that the gravity mediated gaugino masses become
$\lambda^{2s}\Lambda$ which is much smaller than the typical scalar fermion 
mass scale 
$F_S/\Lambda=\lambda^s\Lambda$. This is  because the $S$ field has a non-vanishing 
$U(1)_A$ charge. The gauge mediation is an interesting possibility
to avoid this obstacle, though there is the $\mu$ problem. 
Another obstacle is that comparatively large $D_A$ can induce too large FCNC,
though the stabilizing the moduli makes this issue milder.
We think that this is an interesting and challenging future subject.

The anomalous $U(1)_A$ gauge symmetry plays an important role in solving various
problems in SUSY GUT scenario, for example, the doublet-triplet splitting problem,
the proton stability problem \cite{yukawa6, M1},
unrealistic GUT relation for the Yukawa couplings \cite{yukawa5, yukawa6}, 
the $\mu$ problem \cite{Hempfling, Maekawa-mu}, etc. and in realizing the natural gauge coupling unification.
It is quite impressive that these can be realized 
with a reasonable assumption that all terms which are allowed
by the symmetry of the theory are introduced with $O(1)$ coefficients,
and therefore, all the mass scales can be fixed by the symmetry of the theory.
We had thought that we need an additional sector inducing SUSY breaking, which
is called "Hidden sector" in any SUSY models with the anomalous $U(1)_A$ symmetry.
However, this may not be the case. Adding just
one positively charged field to phenomenologically viable model realizes the 
spontaneous SUSY breaking.
This makes us expect more complete models in which in addition to the previous
advantages of the models with anomalous $U(1)_A$ symmetry, SUSY breaking is also controlled by the anomalous $U(1)_A$ gauge symmetry. 

\section*{Acknowledgement}
S.-G. K., K. S., and N. M. are supported in part by Grants-in-Aid for Scientific 
Research from the Ministry of Education, Culture, Sports, Science 
and Technology of Japan. This work is supported by GCOE Program of Nagoya University
provided by JSPS.



\end{document}